# Mechanisms of information communication and market price movements. The case of SP 500 market.

Inga Ivanova[1] and Grzegorz Rzadkowski[2]

**Abstract**

In this paper we analyze how market prices change in response to information processing among the market participants and how non-linear information dynamics drive market price movement. We analyze historical data of the SP 500 market for the period 1950 - 2025 using the logistic Continuous Wavelet Transformation method. This approach allows us to identify various patterns in market dynamics. These patterns are conceptualized using a new theory of reflexive communication of information in a market consisting of heterogeneous agents who assign meaning to information from different perspectives. This allows us to describe market dynamics and make forecasts of its development using the most general mechanisms of information circulation within the content-free approach.

1. Introduction

A popular approach to understanding the behavior of financial markets involves models that entertain communication between market participants, such as agent-based models (ABMs). ABMs, which take into account the behavioral aspects of investors with different behavioral profiles, such as, e.g. imitation, anti-imitation, and indifference, are used for computer simulation of financial markets (Stefan, & Atman, 2015). The advantage of ABM is that they help mediate the flow of information between individual agents and can be used to study complex phenomena of financial crashes that can't be described by traditional equilibrium based theories. ABMs implicitly assume that information exchange between investors drives market dynamics.

[1] Institute for Statistical Studies and Economics of Knowledge, HSE University, 20 Myasnitskaya St., Moscow, 101000, Russia; inga.iva@mail.ru ; ORCID: 0000-0002-5441-5231.

[2] Department of Finance and Risk Management, Warsaw University of Technology, Narbutta 85, 02-524 Warsaw, Poland e-mail: grzegorz.rzadkowski@pw.edu.pl

Big branch of the literature focuses on machine based time series forecasting. Related model types include traditional Machine Learning (ML) models (e.g. Bahrammirzaee, 2010; Mullainathan, S. and Spiess, 2017; etc.) and emerging Deep Learning (DL) models in the ML field, such as Artificial Neural Networks (ANNs), Recurrent Neural Network (RNNs), Convolutional Neural Network (CNNs) and Deep Multilayer Perceptron (DMLPs) (e.g. Schmidhuber, 2015; Sokolov *et al*., 2020). The commonly observed and difficult to explain facts, such as excess volatility (LeRoy & Porter, 1981; Shiller, 1981), volatility clustering (Mandelbrot, 1997), etc., have led to a growing literature on heterogeneous agent based models (HAMs) (e.g. Hommes, 2006, Chiarella, Dieci, and He, 2009) in which financial market consists of different groups of agents. Interactions between agents can generate complex market dynamics that include both chaos and stability (e.g. Brock, W.A., and Hommes, 1997; Lux, 1995, Bonabeau, 2002). Kaizoji (2004) showed that intermittent chaos in asset price dynamics can be observed in a simple model of financial markets with two groups of agents, which can be explained by heterogeneity in traders' trading strategies.

The behaviour of financial markets during crises can be described using rogue waves (e.g. Jenks, 2020). This type of wave is analytically described in the nonlinear option pricing model (Yan, 2010). These solutions can be used to describe possible mechanisms of the rogue wave phenomenon in financial markets. Rogue waves can be found in Korteweg de-Vries (KdV) systems if real non-integrable effects, higher order nonlinearity, and nonlinear diffusion are taken into account (Lou & Lin, 2018). Dhesi & Ausloos (2016) observed a kink-like effect resembling a soliton behavior when studying the behavior of an agent reacting to time-dependent news on log returns in an Irrational Fractional Brownian Motion model. They also asked what is the differential equation whose solution describes this effect?

The disadvantage of the computational ABM approach is that, while representing the result of collective interaction of investors, they, being the down-top approach, lack in our understanding the global aspects of market dynamics that are provided by top-down theories. The advantage of the analytical approach in comparison with the numerical one is that in analytical models it is possible to trace the properties of market time series to the behavior of investors.

A common assumption is that asset prices are determined by investors' reaction to information. However they overlook the mechanisms by which investors process this information. One can ask questions: 1) how investors perceive information, 2) does the same information have the same meaning for different groups of investors, 3) does the way investors process information matter more than the information itself? The motivation of the paper is to find an answer to these questions.

The contribution of this paper is twofold: 1) in a narrow sense it presents a model of asset price dynamics which can shed light on the mechanisms that govern the formation of asset price time series and, in some cases, can predict future price movement; 2) in a broad sense, it is a step forward in developing a theory of meaning that moves this theory into broader practical areas.

The first research objective of this paper is to test the applicability of the general concept of information and meaning to describe the dynamics of market prices. The second research objective is to provide a quantitative description of the movement of market prices based on the evolutionary dynamics of investors' expectations. We show that this non-linear dynamics can be captured by a nonlinear evolutionary differential equation whose solutions yield recognizable patterns. These patterns can be used to predict the future movement of market prices. This allows more accurate predictions of future price movements of market assets and market crashes.

The second aspect of this paper concerns the modelling of financial time series. The dynamics of asset prices have been the subject of the debate among academics and practitioners.

Can the description of markets within the framework of the reflexive information and communication approach adequately reflect the observed dynamics?

In this paper we provide explanations for the above questions in the form of content-free information theory that describes the market in terms of systemness.

## 2. Concept

The study is based on systems theory where market is considered as a system comprised of interacting heterogeneous groups of investors whose interaction is grounded on meaningful processing the information. Meanings supplied to information drive investors' decisions and generate market price movement. In some sense this approach is close to that used in opinion dynamics models (e.g. Zha *et al.*, 2020; Granha et al., 2022). Here the term "heterogeneous" implies that investors from different groups have different meaning generation mechanisms to process information so that the same information is supplied with different meanings. When information is received it should first be processed, i.e. supplied with meaning. However information can be processed differently by different groups of investors, which provide differing criteria with which to supply information with meaning. These criteria can be specified as selection environments in terms of specific coding rules (or sets of communication codes). Coding rules drive latent structures which organize different meanings into structural components. *"Meanings originate from communications and feedback on communications. When selections can operate upon one another, a complex and potentially non-linear dynamics is generated"* (Leydesdorff, 2021, at p.15).

Meanings produce expectations about possible system states which are generated with respect to future moments of time. Expectations operate as a feedback on the current state (i.e. against the arrow of time). In other words the system simultaneously entertains it's past, present and future states, this accords with Bachelier's observation that "*past, present and even discounted future events are reflected in market price*" (Bachelier, 1900). Expectations provide a source of additional options for possible future system states that are available but have not yet been realized. The more options possessed by the system the greater is the likelihood that the system will deviate from the previous state in the process of autocatalytic self-organization. The measure for additional options is provided by redundancy which is defined as the complement of information to the maximum informational content (Brooks & Wiley, 1986). Redundancy evolution can eventually generate non-linear dynamics in investors' expectations. Quantitative approach to measuring redundancy is formulated within the Quantitative Theory of Meaning which builds upon the seminal works of Loet Leydesdorff on the dynamics of expectations and meaning in inter-human communications (Leydesdorff, 2008; Leydesdorff & Dubois, 2004; Leydesdorff, Dolfsma, Van der Panne, 2006; Leydesdorff & Franse, 2009; Leydesdorff & Ivanova, 2014; Leydesdorff, Petersen & Ivanova, 2017). The theory also incorporates the conceptual framework of the Triple Helix model of university-industry-government relations (Etzkowitz & Leydesdorff, 1995, 1998) and its mathematical formulation (Ivanova & Leydesdorff, 2014a, 2014b).

It is implied that systems' dynamics is driven by reflexive communication of information. Mechanisms of information and meaning processing are different. Whereas information can be communicated via network of relations meanings are provided from different positions (Burt,

1982). Meaning cannot be communicated but only shared when the positions overlap. Processing of meaning can enlarge or reduce the number of options which can be measured as redundancy. Calculus of redundancy is complementary to calculus of information.

Shannon (1948) defined information as probabilistic entropy: $H = -\sum_i p_i \log p_i$ which is always positive and adds to the uncertainty. One can consider two overlapping distributions with information contents $H_1$ and $H_2$ (Figure 1):

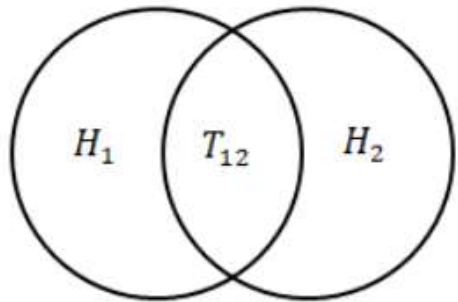

**Figure 1**: Set-theoretical representation of two overlapping distributions with informational contents $H_1$ and $H_2$

Total distribution is a sum of two distributions minus overlapping area, since it is counted twice:

$$H_{12} = H_1 + H_2 - T_{12} \qquad (1)$$

Overlapping area relates to mutual, or configurational (McGill, 1954) information ($T_{12}$). The aggregate distribution $H_{12}$ is:

$$T_{12} = H_1 + H_2 - H_{12} \qquad (2)$$

Analogously mutual (or configurational) information in three dimensions $T_{123}$ (e.g., Abramson, 1963) is (Figure 2):

$$T_{123} = H_1 + H_2 + H_3 - H_{12} - H_{13} - H_{23} + H_{123} \qquad (3)$$

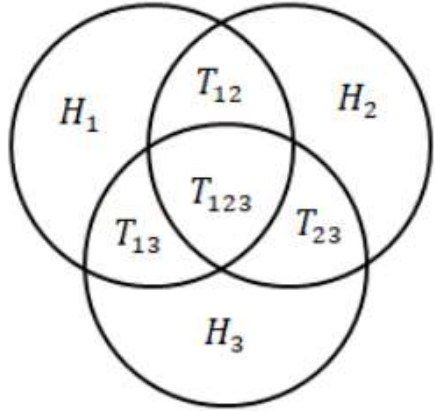

**Figure 2:** Set-theoretical representation of three overlapping distributions with informational contents $H_1$, $H_1$ and $H_3$

Configurational information $T_{123}$ is no longer Shannon-type information since it can be negative. The sign alters with each newly added distribution (e.g. Krippendorff, 2009). Technically the problem of sign change can be solved by introducing "positive overlapping" (Leydesdorff & Ivanova, 2014). This time one does not correct for overlap which is counted twice and therefore redundant, but assume other mechanisms with which two distributions influence one another. These mechanisms are different from relational exchange of information and lead to increase of redundancy so that overlapping area is added instead of being subtracted. In formula format:

$$H_{12} = H_1 + H_2 + R_{12} \qquad (4)$$

It follows that $R_{12}$ is negative ($R_{12} = -T_{12}$) and is hence a redundancy - reduction of uncertainty[3], analogously: $R_{123} = T_{123}$[4]. That is measuring configurational information in three (or more) dimensions one measures not Shannon-type information but mutual redundancy. Since this measure provides a negative amount of information it can be considered as an indicator of synergy among three sources of variance (Leydesdorff, 2008).

---

[3] Redundancy is defined as a fraction of uncertainty that is not used for the uncertainty $H$ that prevails: $R = 1 - \frac{H}{H_{max}} = \frac{H_{max}-H}{H_{max}}$. $H_{max}$ – maximum possible entropy. Adding new options increases $H_{max}$ and redundancy.
[4] In fact, a more general condition must be satisfied: $R_{12} = -\alpha T_{12}$

When a system comprises three (or more) groups of agents each third group disturbs interaction between the other two. This mechanism is known as "triadic closure" and drives the system evolutionary dynamics (Granovetter, 1973; Bianconi *et al.*, 2014, de Nooy & Leydesdorff, 2015). Triads can be transitive or cyclic (Batagelj *et al.*, 2014). Two cycles may emerge – positive (autocatalitic) and negative (stabilizing) ones Autocataitic cycle reinforce the change from previous system state (the system self-organises) while stabilising cycle keeps the system from transformation. The balance between stabilization and self-organization can captured by the formula *P* and *Q* (Ivanova & Leydesdorff, 2014a, 2014b):

$$R \sim P^2 - Q^2 \tag{5}$$

Here *P* and *Q* - three dimensional vectors. The first term in Eq. 5 is related to historical realization (which adds to positive entropy) and the second term - to self-organization and augments negative entropy. Historical realization relates to historically realized options which are generated via recursive mode and self-organization bears on new, not yet realized options, generated via incursive mode[5]. The trade-off between historical realization and self-organization (Eq. 5) leads to non-linear evolutionary equations for redundancy (Ivanova and Rzadkowski, 2025)[6]:

$$4R_T - 2\alpha R R_X + R_{XXX} + C_1 = 0 \tag{6}$$

which is generalization of Korteveg-de Vries equation (KdV):

---

[5] Recursive systems use their past states to modulate the present ones, incursive (or anticipatory) systems employ possible future states to shape their present states (e.g. Rosen, 1985; Dubois, 1998; Leydesdorff & Dubois, 2004)
[6] More precisely, it is a family of equations determined by the scale factor $\alpha$.

$$u_T - 6uu_X + u_{XXX} = 0$$

The important feature of KdV equation is the existence of soliton solutions:

$$u(X,T) = -2k^2 \cosh^{-2}(kX - 4k^3 T)$$

Initial arbitrary impulse develops with time into one of few solitons (if any) and oscillatory waves. For the chain of solitons originated from the same initial impulse the ratio of soliton amplitudes to the time shifts is constant

$$A_i / T_i = const \tag{7}$$

In other words, the chain of solitons forms a linear trend.

## 3. Method

We analyze S&P 500 market data using the method of wavelet transform. This method presents an effective technique for compression/decompression temporal signals. The core advantage of wavelet transform is data compression so that signal is represented by a smaller amount of information – just by a few wavelets which comprise statistically concentrated information about the signal. There are many different types of wavelets (e.g. Mexican hat, Morlet, Daubechies, Gabor, Haar, etc. wavelets) used for signal compression. However the results of wavelet transform can be hardly conceptualized in terms of signal and signal generating system relating properties. To overcome this obstacle we implement specific form of transform - logistic CWT (cf. Rzadkowski and Figlia, 2021, Ivanova and Rzadkowski, 2025) - suggested by explicit solution of Eq. 6 which corresponds to the first derivative of the logistic function.

A mother wavelet (Daubechies, 1992, p.24) is an integrable function $\psi \in L^1(\mathbb{R})$, which satisfies the following admissibility condition:

$$2\pi \int_{-\infty}^{\infty} |\xi|^{-1} |\hat{\psi}(\xi)|^2 d\xi < \infty \qquad (8)$$

where $\hat{\psi}(\xi)$ is the Fourier transform of $\psi$

$$\hat{\psi}(\xi) = \frac{1}{\sqrt{2\pi}} \int_{-\infty}^{\infty} \psi(t) e^{-i\xi t} dt$$

Moreover, we assume that $\psi$ is square integrable, with the norm:

$$\|\psi\| = \|\psi\|_{L^2} = \left( \int_{-\infty}^{\infty} |\psi(t)|^2 dt \right)^{1/2}$$

By dilating and translating of the mother wavelet one obtains a family of wavelets (so called children wavelets)

$$\psi^{\alpha,\beta}(t) = \frac{1}{\sqrt{\alpha}} \psi\left(\frac{t-\beta}{\alpha}\right)$$

where $\alpha, \beta \in \mathbb{R}$, $\alpha > 0$. It is seen that $\|\psi^{\alpha,\beta}\| = \|\psi\|$. Usually the mother wavelet is normalized, $\|\psi\| = 1$. Continuous Wavelet Transform (CWT) of a function (signal) $f \in L^2(\mathbb{R})$ at a point $(\alpha, \beta)$ is the inner product of $f$ and the wavelet $\psi^{\alpha,\beta}$

$$(T^{wav} f)(\alpha, \beta) = \langle f, \psi^{\alpha,\beta} \rangle = \int_{-\infty}^{\infty} f(t) \psi^{\alpha,\beta}(t) dt \qquad (9)$$

where $\alpha, \beta$ are parameters of the wavelet family. For each fixed value of $\alpha$, formula (9) is the convolution of two functions, the signal and the wavelet. In Matlab, applying CWT to a function produces a graph called a scalogram, in which the CWT values are given as colors.

Starting from the logistic function $x(t) = \frac{1}{1+e^{-t}}$, which is also the cumulative distribution function of the standard logistic distribution, we use, for modelling, the second-order normalized logistic mother wavelet (Rzadkowski and Figlia, 2021, Ivanova and Rzadkowski, 2025)

$$\psi_2(t) = \sqrt{30}\, x''(t) = \frac{\sqrt{30}(e^{-2t} - e^{-t})}{(1+e^{-t})^3} \qquad (10)$$

and $\psi_2^{\alpha,\beta}(t) = \psi_2\left(\frac{t-\beta}{\alpha}\right)$.

Ivanova and Rzadkowski (2025) proved that the CWT (9) with logistic wavelets $\psi_2^{\alpha,\beta}$

$$(T^{wav}f'')(\alpha,\beta) = \langle f'', \psi_2^{\alpha,\beta} \rangle = \int_{-\infty}^{\infty} f''(t)\psi_2^{\alpha,\beta}(t)dt \qquad (11)$$

applied to the second derivative $f''(t)$ of the function

$$f(t) = c + dt + \frac{y_{sat}}{1+\exp\left(-\frac{t-b}{a}\right)}$$

takes maximum (for $y_{sat} > 0$) or minimum (for $y_{sat} < 0$) when $\alpha = a$ and $\beta = b$.

In this paper, we examine the monthly historical data of the S&P 500 stock market index for the period from July 1982 to April 2025[7]. This gives a time series consisting of 514 observations. Let us denote these observations by $x_n$ ($n = 1,2,\ldots,514$; $n = 1$ for 01.07.1982, $n = 514$ for 01.04.2025) and the time series by $(x_n)$.

Using a multilogistic function with an additional linear component

$$y(t) = c + dt + \sum_{i=1}^{k} \frac{y_{i,sat}}{1+\exp\left(-\frac{t-b_i}{a_i}\right)} \qquad (12)$$

where $c, d$ – real numbers, $y_{i,sat}$ – saturation level, $b_i$ – shift and $a_i$ – slope coefficient of the $i$th logistic function, we formally approximate the total data $y_n = \sum_{i=1}^{n} x_i$ and in order to approximate the differential data $(x_n)$, i.e., the monthly data of the S&P 500 index, we use the derivative of this function

$$y'(t) = d + \sum_{i=1}^{k} \frac{y_{i,sat}\exp\left(-\frac{t-b_i}{a_i}\right)}{a_i\left(1+\exp\left(-\frac{t-b_i}{a_i}\right)\right)^2} = d + \sum_{i=1}^{k} \frac{y_{i,sat}}{4a_i} \cosh^{-2}\left(\frac{t-b_i}{2a_i}\right) \qquad (13)$$

So, in order to find solitary waves in the time series $(x_n)$ we apply CWT (11) to its first differences

$$\Delta^1 x_n = x_n - x_{n-1}$$

In Matlab, the cwt command is applied for this purpose. Because the signal is discrete, then instead of the integral, cwt command uses its Riemann sum. As a result one obtains a scalogram in which the value of the integral CWT is given as a color. For every pair of parameters $(\alpha, \beta)$, the value of CWT (11) is written in color (the scale value-color is seen on the right-hand side of

---

the scalogram). In the scalogram we are looking for points at which the CWT (color) takes locally maximal or minimal value. Such points give us parameters $(a, b)$ of consecutive waves. From the value of CWT (color) we calculate the third parameter $y_{sat}$.
Algorithmic formalization of logistic CWT and the complete Matlab code can be found in Ivanova and Rzadkowski (2025).

In cases where there is no clear maximum or minimum in some scalogram region, we use local parameter optimization to better fit the approximating function to the data. In case of overlapping waves we can remove higher intensity wave from the time series and repeat the CWT. This step can be taken several times to make lower intensity waves visible on CWT scalogram - a region in a *time-scale* plane where the color indicates statistical significance of CWT coefficients.

## 4. Results

We decompose the historical monthly SP 500 inflation adjusted data for the period 1982-2025[8] into a sum of 22 waves. The results are presented in Figs.3-6. SP 500 trend dynamics follows the dynamics of innovation system synergy in the framework of the Triple Helix model. Wavelet decomposition reveals hidden trend structure comprised of positive and negative sequences of solitary waves. Using the method described in Section 3, we obtained scalograms for the parameter $\alpha$ in the interval $(1, 30)$ (Fig. 3) and also in the interval $(1, 120)$ (Fig. 4) to determine large carrier waves. Fig. 4 shows two large waves A, B, the sum of which, after estimating the parameters, we have taken as one carrier wave (Figs. 10, 11). Then we specified 10 relatively large positive waves (Fig. 5), and 10 negative waves (Fig. 6). We omitted the remaining waves because we considered that the obtained fit to the data ($R^2 = 0.9939$) is already sufficiently accurate. The estimated parameters of all 22 waves are listed in Table 1. The approximation function (13) with parameters from Table 1 and with the additional parameter $d = 13.9$ is shown in Fig. 7.

---

[8] retrieved from: https://www.macrotrends.net/2324/sp-500-historical-chart-data

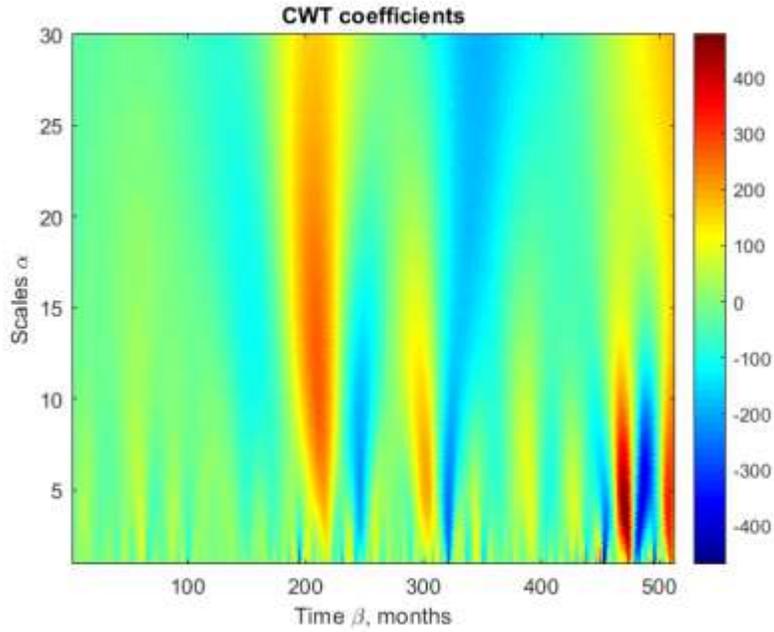

**Figure 3**: Scalogram, 514 months ('1'-01.07.1982, '514' -01.04.2025)

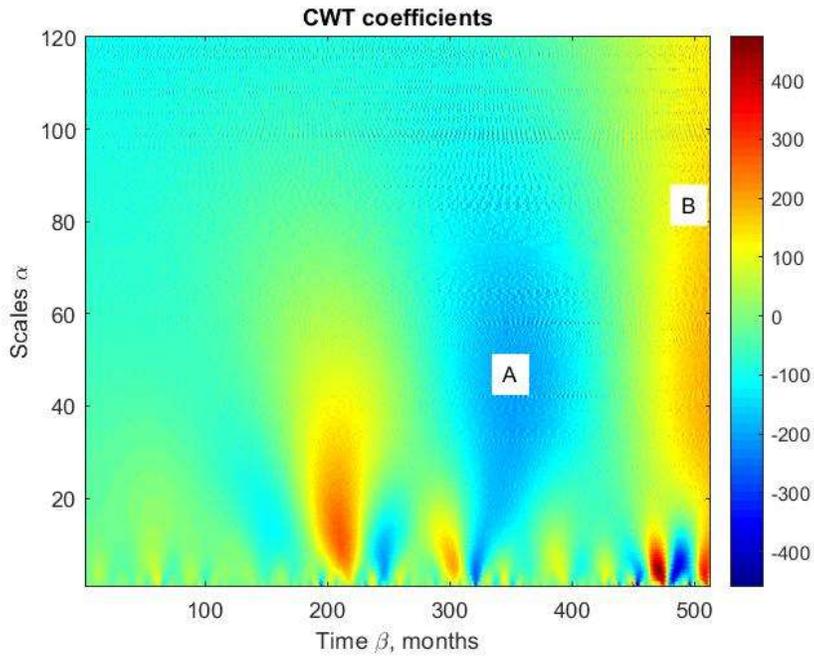

**Figure 4**: Carrier waves – A (negative) and B (positive)

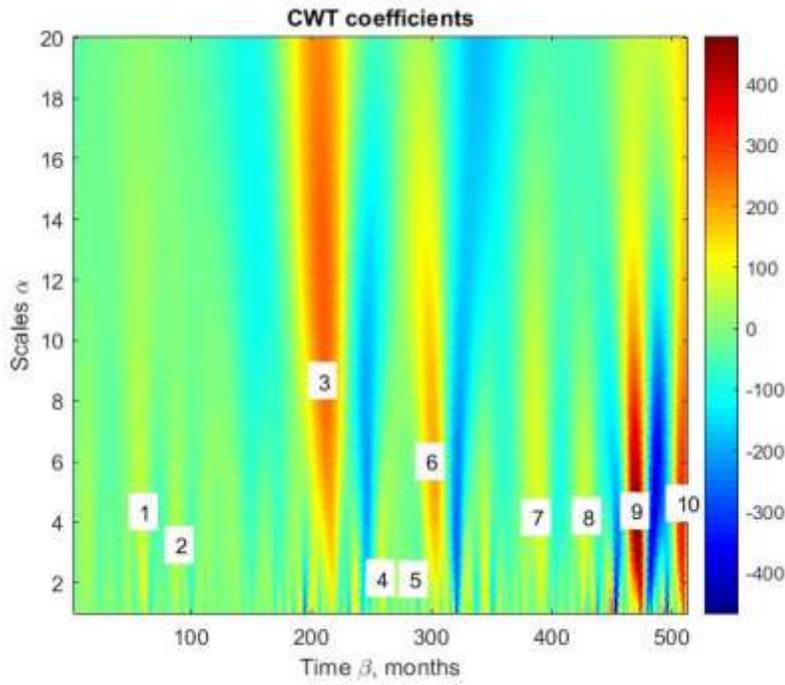

**Figure 5**: Positive waves 1-10

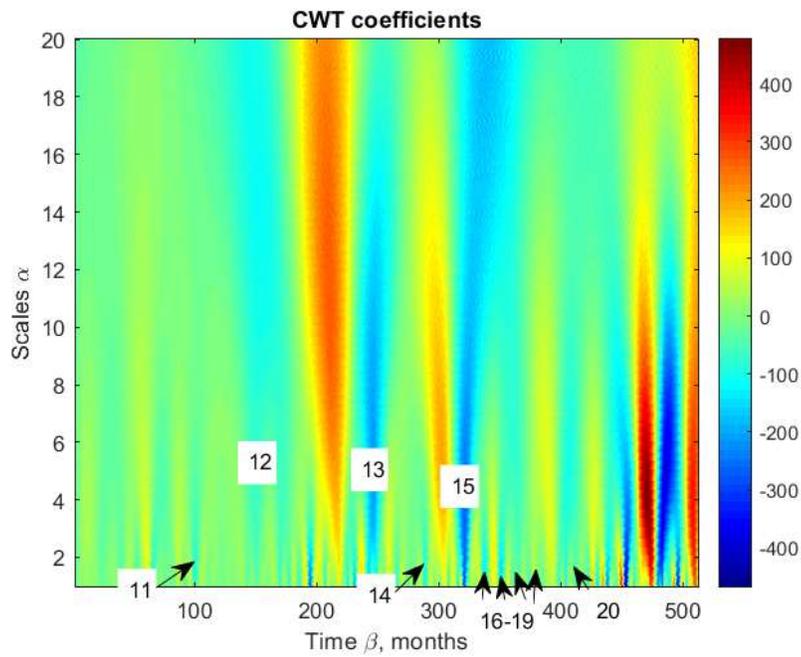

**Figure 6**: Negative waves 11-20

**Table 1**: Wave parameters (b-time, a-scale, $y_{sat}$-saturation level) A, B -carrier waves

| wave | a | b | ysat | wave | a | b | ysat |
|------|------|------|---------|------|------|-----|--------|
| A | 40.7 | 354 | -201951 | 10 | 4.5 | 509 | 33808 |
| B | 139.5 | 511 | 2377300 | 11 | 1.5 | 101 | -500 |
| 1 | 4.3 | 60 | 3419 | 12 | 3 | 151 | -1700 |
| 2 | 3.4 | 88 | 1650 | 13 | 4.4 | 246 | -5042 |
| 3 | 11.6 | 209 | 57083 | 14 | 1.9 | 289 | -1096 |
| 4 | 1.8 | 259 | 1651 | 15 | 5.4 | 319 | -15831 |
| 5 | 2 | 270 | 1368 | 16 | 1.2 | 321 | -608 |
| 6 | 10 | 303 | 28921 | 17 | 1.9 | 338 | -1600 |
| 7 | 4.9 | 390 | 5600 | 18 | 1.5 | 352 | -1300 |
| 8 | 3.6 | 427 | 3460 | 19 | 1.7 | 365 | -700 |
| 9 | 4.4 | 471 | 23238 | 20 | 4.6 | 406 | -4500 |

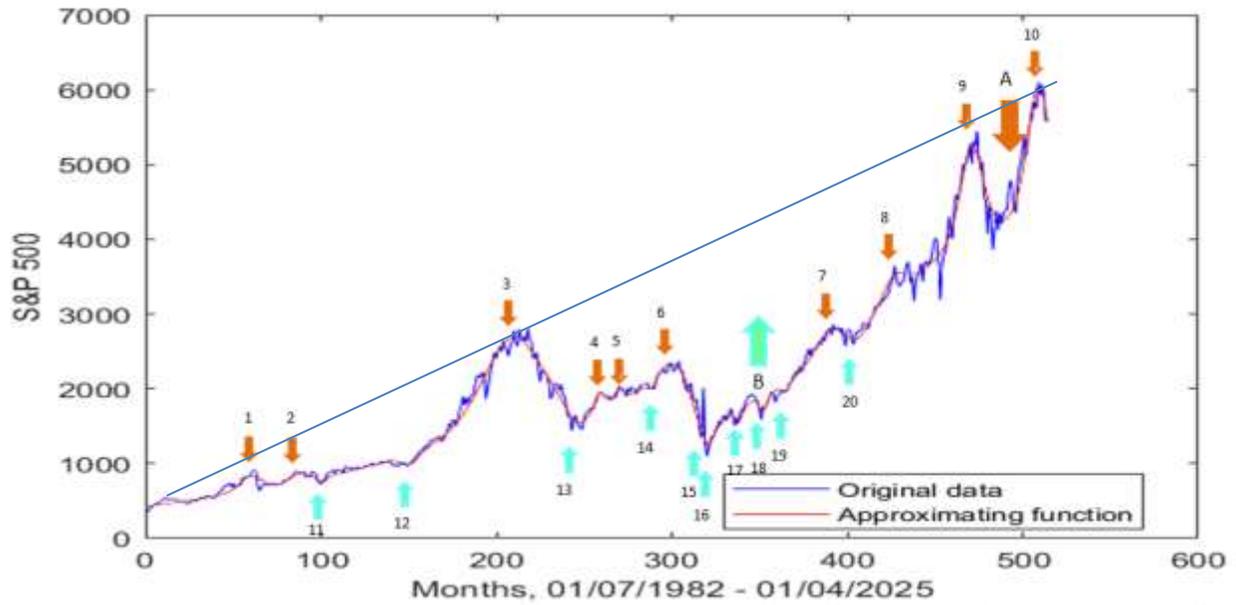

**Figure 7**: Approximation by the first derivative of a multilogistic function (sum of 22 logistic functions) ($R^2$ =0.99390868, RMSE=103.24)

The clear appearance of the smaller positive (downward arrows) and negative (upward arrows) functions and carrier waves is shown in Figures 8 -11. The straight trend lines in Figures 8 and 9 indicate wave sequences characterized by a ratio 7.

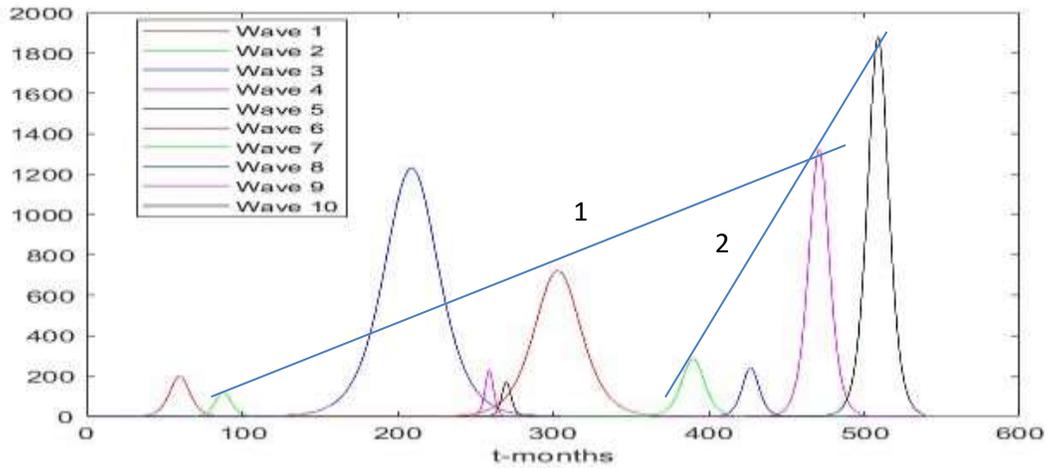

**Figure 8**: Positive waves (1-10)

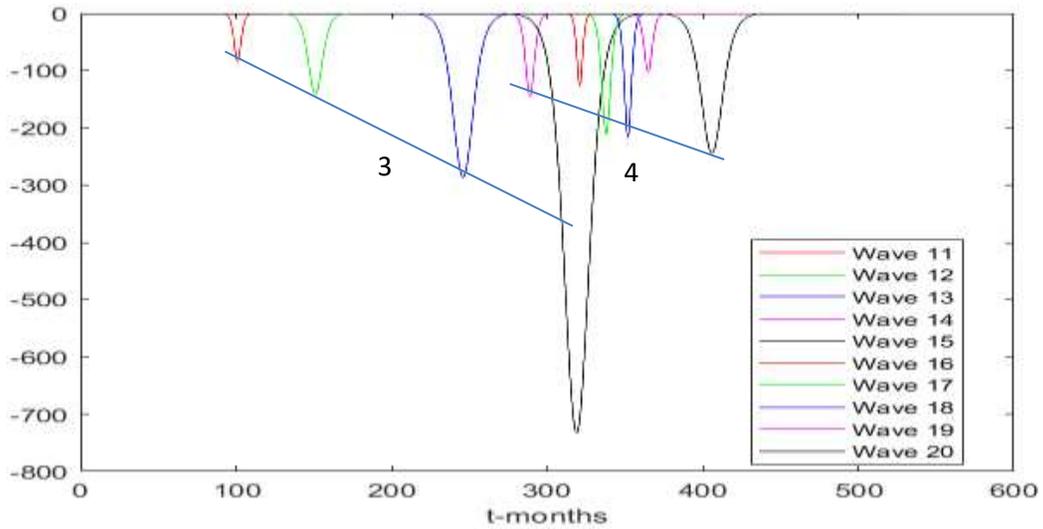

**Figure 9**: Negative waves (11-20)

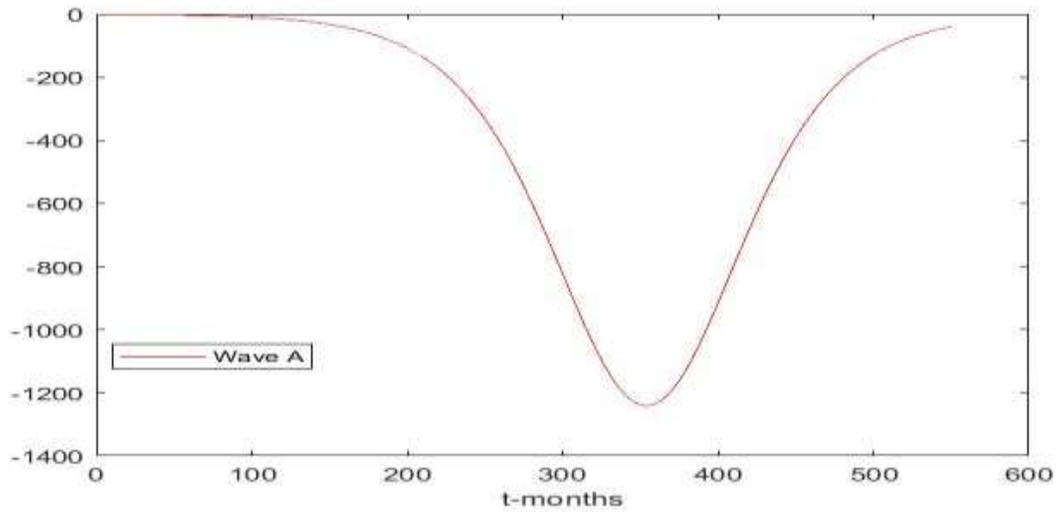

**Figure 10**: Carrier wave A

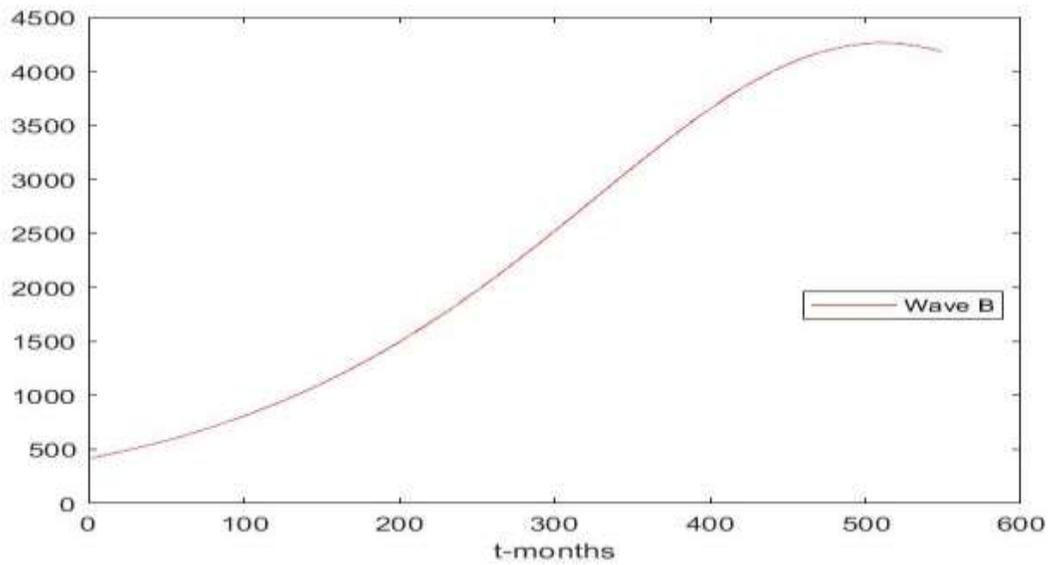

**Figure 11**: Carrier wave B

We conclude that the dynamics of the SP 500 trend can be adequately reflected by the multi-soliton decomposition. In turn, this multi-soliton dynamics is predicted within the Triple Helix model. Wavelet decomposition reveals the hidden structure of the trend, consisting of positive and negative sequences of single waves which can be considered in terms of market systemness estimated as synergy. The Wave values indicate the amount of (positive and negative) uncertainty generated. Positive and negative waves can be attributed to two competing processes which are present: the historical generation of variations and interaction between selection environments. While the historical generation of variations increases the overall uncertainty of the market, the interaction between selection environments leads to generation of new options, which leads to a decrease in uncertainty (Petersen, Rotolo, & Leydesdorff, 2016). The trade-off between these processes determines the sign of the synergy. Positive synergy indicates that Shannon-type information generation is prevalent, system uncertainty is increasing and market is forming an upward trend. Negative synergy implies that non-linear generation of redundancy in loops of communication prevails and market uncertainty decreases leading to a downward trend. The Efficient Market Hypothesis (EMH), formalized by Fama (1970), states that asset prices fully reflect all available information, making markets unpredictable. Entropy-based measures provide a robust way to test this hypothesis by quantifying deviations from randomness. Lower entropy indicates predictability, often observed during financial crises when market behavior becomes more deterministic due to panic-driven trading or regulatory interventions (Papla & Siedlecki, 2024).

A sharp decrease in uncertainty can be defined as a market crash. A large-scale and prolonged crash can be considered a financial crisis. It is widely believed that there is an interaction between financial cycles and the broader macroeconomic business cycle. Economic recessions can trigger financial crises and vice versa. The economic consequences of financial crises are usually severe and long-lasting, with output, employment, and investment often experiencing deeper and longer declines than in normal recessions (Reinhart & Rogoff, 2009; Jordà *et al*., 2013). Although it has traditionally been argued that financial cycles occur with lower frequency and higher amplitude than business cycles, recent research has challenged this view. Cagliarini and Price (2017) find that financial and macroeconomic cycles often exhibit overlapping lengths and significant synchronization. Chen and Svirydzenka (2021) demonstrate that financial cycles—measured through credit, equity, and property prices—typically span medium-term frequencies (around six years), with equity prices and output gaps serving as reliable early warning indicators in advanced economies. Komorowski (2017) extends this discussion by proposing a life cycle model of economic crises, emphasizing the sequential nature of crisis phases and the tendency for new "normals" to emerge. In this model, the crisis unfolds through stages, from the slow accumulation of systemic risks to the sudden burst of speculative bubbles, widespread contagion, and ultimately the establishment of irreversible changes in economic structure and behavior. Such a model highlights the importance of understanding crisis "transmission" dynamics, as seemingly insulated or resilient economies may experience delayed, but ultimately severe, impacts as global transmission progresses.

The negative peaks 11-20 in Fig.9 are associated with financial market crises, such as US Savings and Loan crisis (wave 11), The Mexican economic crises (wave 12), The Early 2000s recession (wave 13), the 2008 financial crisis (wave 15), the Black Monday 2011 (wave 18), the January 2016 global meltdown (wave 20) etc. Although the peaks corresponding to 2020 (Covid 19 pandemic) and 2022, indicated by the large blue color in scalogram in Fig. 3, located before the large positive waves 9 and 10, are missing. The large CWT value (color) is created by the decrease in soliton 9 and the subsequent increase in soliton 10. This explains the creation of this color.

The market can operate in different modes at the same time, some modes are represented by only one wave, and some generate a sequence of waves. There is an opinion that the subsequent crisis can be a logical continuation of the previous one (trends 1, 2 in Fig. 8 and trends 3, 4 in Fig. 9). The trend existence gives an opportunity for forecasting the trend reversal. The appearance of the third wave, as in trend 2 (Fig. 8), suggests an inevitable trend change. The same applies to negative trends. The successive formation of negative waves 11, 12, 13 predicts an upswing, indicated by waves 4, 5, 6. The same for the negative trend formed by waves 14, 17, 18, 20, the end of which is marked by a sharp rise in the market (waves 8, 9, 10). The overall picture of further market dynamics is negativeю This is confirmed by the fact that the positive carrier wave B (Fig. 11) has passed its turning point, and the negative carrier wave A (Fig. 10) is at its lower end. The formation of the trend can be consistently traced. If, for example, we take half of the time series (e.g. 0-300), we will get the same positive and negative waves. However, at the end of the region, differences may arise. When the wavelet moves along the signal and reaches the end of the region (i.e. 300 on the left), the integral (Eq. 14) will cover only one half of it. This is where these differences come from.

5. **Discussion**

In present paper we analyze market dynamics using information theory. The application of information theory to financial markets has emerged as a transformative framework for analyzing market efficiency, risk dynamics, and price behavior. Originating from Claude Shannon's foundational work on communication theory (Shannon, 1948), information theory provides quantitative tools to measure uncertainty, complexity, and information flow—concepts that are inherently relevant to financial systems. Over the past decades, researchers have increasingly adopted entropy-based measures, fractal analysis, and related methodologies to uncover hidden patterns in financial data, assess market efficiency, and optimize trading strategies

Shannon's entropy, in particular, has proven instrumental in evaluating market behavior. Studies such as Carranza *et al.* (2023) demonstrate its effectiveness in quantifying complexity in financial time series, while Papla and Siedlecki (2024) show that entropy levels fluctuate during economic crises, reflecting changes in market efficiency due to external interventions. Beyond traditional entropy, other information-theoretic approaches, such as Tsallis entropy and approximate entropy, have been employed to differentiate between short-term market crashes and prolonged crises (Gençay & Gradojevic, 2017). Recent advancements extend these applications to portfolio optimization, where models like the Mean-Absolute Deviation-Entropy (MAD-entropy) enhance diversification by minimizing unsystematic risk (Sinon & Mba, 2024). However, critiques remain regarding the limitations of entropy measures in highly nonstationary markets, suggesting the need for hybrid methodologies that combine traditional statistical tools (Papla & Siedlecki, 2024).

The application of Shannon entropy extends beyond high-frequency data to broader market analyses. Cho and Kim (2022) introduced attention entropy, a variant of Shannon entropy, to assess the time-varying efficiency of 27 global financial markets. Their study revealed hierarchical clustering based on long-term efficiency trends, with short-term instability driven by systemic risks and global events such as the COVID-19 pandemic. This underscores entropy's utility in capturing both localized and systemic inefficiencies.

While Shannon entropy measures uncertainty, Transfer Entropy (TE) quantifies directional information flow between financial time series, offering insights into market interdependencies. Marschinski and Kantz (2002) pioneered the use of TE to analyze information transfer between the Dow Jones and DAX indices, demonstrating its superiority over linear correlation in detecting nonlinear dependencies. Their work introduced the Effective Transfer Entropy (ETE) estimator to mitigate finite-sample biases, revealing a stronger directional influence from the U.S. market to Germany, reflecting economic dominance.

Building on this, Liu et al. (2020) expanded TE applications to identify distinct market regimes—return-driven, news-driven, and mixed—by modeling financial markets as a bivariate system of news sentiment and returns. Their findings showed that during crises (e.g., the 2008 liquidity crisis and Eurozone debt crisis), markets shifted to a news-driven regime, highlighting sentiment's role in periods of uncertainty. This work validated entropy measures in regime identification and provided insights into adaptive market efficiency.

Mutual Information (MI), another key information-theoretic tool, measures the general dependence between variables without assuming linearity. Korbel et al. (2019) employed Rényi transfer entropy (RTE) to analyze information transfer in financial networks, finding that banking and real estate sectors act as primary hubs for risk propagation during crises. Their study revealed that nonlinear interactions dominate during extreme events, leading to negative RTE values in developed markets like the NYSE, indicating heightened complexity.

Similarly, Benedetto et al. (2021) applied MI to study interactions between commodity and financial asset prices, uncovering nonlinear dependencies that traditional correlation-based models miss. This approach provides a more nuanced understanding of cross-market information flows, particularly relevant for portfolio diversification and systemic risk assessment.

Information theory has also advanced the study of long-range dependence (LRD) in financial markets. Murialdo et al. (2020) proposed the Moving Average Cluster Entropy (MACE) framework to detect LRD, revealing persistent multi-scale volatility patterns. Ponta et al. (2021) extended this by developing an entropy-based measure for horizon dependence, confirming that financial markets frequently deviate from EMH, displaying statistically significant autocorrelations across time horizons. These studies challenge traditional econometric models by demonstrating that entropy-based methods better capture market inefficiencies and predictability.

The globalization of finance has further complicated market dynamics, as cross-border capital flows and interconnected markets heighten contagion risks. Integrating alternative data sources (e.g., social media sentiment) may refine entropy-based measures of market behavior. The growing complexity of financial markets calls for continued innovation in entropy methodologies to capture emerging patterns and risks.

Sufi & Taylor (2022), drawing from behavioral finance, highlight how over-optimism and extrapolation of past trends can foster periods of excessive risk-taking and borrowing, contributing to credit and asset price booms that ultimately end in busts. This perspective is further complicated by debates on whether endogenous randomness within markets, or the actions of heterogeneous individuals and institutions, primarily drives such cyclical excess (Brunnermeier & Oehmke, 2013; Aikman *et al*., 2014; Dávila & Korinek, 2017). The literature also recognizes that even rational models of market interaction can produce credit booms and asset price bubbles if incentives are misaligned, if government backstops are anticipated, or if there are pecuniary externalities and strategic complementarities at play. This fusion of rational and behavioral insights has enhanced theoretical models and informed improved empirical methodologies for crisis forecasting.

In our approach, we combine concepts and methods from information theory, sociology, physics, mathematics, and computational modeling to provide a broad approach to analyzing financial markets. Instead of analyzing specific forms of interaction between investors and the macroeconomic situation, we draw on the general theory of reflexive communications. The typical dynamics of a complex system predicted by the theory are supported by observed financial market phenomena which creates ground for further discussions. On the one hand, market movement is determined not only by the investors' preferences, but also by external objectively existing circumstances determined by economic, political, scientific, technological and other factors. On the other hand, one can ask the question: is the emergence of new scientific discoveries, inventions, innovations, new technologies, economic structures a consequence of the fundamental laws of information circulation?

## 6. Conclusion

This paper complements the studies devoted to theory application to analysis of complex systems of different origins describing Covid-19 infectious disease spread, rumors propagation, patent dynamics, and FOREX market (Ivanova, 2022; Ivanova and Rzadkowski, 2025; (Ivanova, Rzadkowski, and Leydesdorff, 2025).

Meaning in social communications is processed via specific sets of communication codes which span horizons of meaning acting as selection and coordination mechanisms. It is based on expectations and incurred on events against the arrow of time. Expectations can be measured as redundancy (i.e. additional options) with help of Shannon's entropy information theory. Redundancy dynamics is captured by non-linear evolutionary equation which has soliton solutions.

The market is considered as an ecosystem with bi- and trilateral relations among the agents. In this respect mechanisms that drive market evolution are similar to mechanisms of the TH model of innovations. Asset price dynamics can be analyzed from an information theory perspective taking into account the relationships between information processing and meaning generation. Agents represent three groups of investors with preferences to hold long and short positions, or temporally abstain from active actions which can be considered distributions spanning the network of relations. Information is communicated via the network of relations. There are other dynamics on the top of this network of relations. Different agents use different communication codes, reflecting their preferences, to provide meaning to the information. Codes can be considered the eigenvectors in a vector-space (von Foerster, 1960) and structure the communications as selection environments. Communicated information is supplied with different meaning by different agents. Meaning is provided from the perspective of hindsight. Meanings cannot be communicated, as in case of information, but only shared. Providing information with meaning increases the number of options (redundancy). This mechanism can be considered probabilistically using Shannon's equations (Shannon, 1948). The generation of options (redundancy) is crucial for system change. The trade-off between the evolutionary generation of redundancy and the historical variation providing uncertainty can be measured as negative and positive information, respectively. The dynamics of information, meaning, and redundancy can be evaluated empirically using the sign of mutual information as an indicator. When the dynamics of expectations, generating redundancies, prevail over the historical construction generating entropy, mutual redundancy is negatively signed because the relative uncertainty is reduced by increasing the redundancy. The balance between redundancy and entropy can be mapped in terms of two vectors ($P$ and $Q$) which can also be understood in terms of the generation versus reduction of uncertainty in the communication that results from interactions among the three (bi-lateral) communication channels. Eventually non-linear mechanisms in redundancy evolution may prevail which gives rise to the predictable behavior of market price evolution. This does not imply that market price movement is totally predictable, but in certain periods plausible assessment of price development can be made.

Information theory provides a powerful lens for dissecting financial markets, capturing nonlinear interactions, and identifying regime shifts. From Shannon entropy's role in assessing market efficiency to transfer entropy's application in detecting information flows, these methodologies offer deeper insights than traditional linear models. Empirical studies demonstrate their effectiveness in risk measurement, portfolio optimization, and systemic risk analysis. However, challenges remain in adapting these methods to nonstationary environments and large-scale datasets. Future research should focus on hybrid approaches and real-time applications, ensuring that entropy-based models remain at the forefront of financial market analysis.

Employing the dynamics of expectations and meaning makes it possible to make a bridge between EMH, behavioral economics and technical analysis. The market accepts all available information but this information is supplied with different meanings by different agents and triggers different actions, which can sometimes be interpreted by an external observer as non-rational behavior. The interaction among agents can at times generate persistent tendencies. This suggests that different approaches to describing market dynamics can be employed from a unified point of view. The subject of future studies comprises the application of the present approach to some other problems and datasets with a varying numbers of features.

The paper findings can also be useful for market practitioners in their daily activities. This conceptual and modeling approach may be widely used by researchers seeking to study the evolution of complex adaptive systems consisting of communicating heterogeneous agents.

# References

Abramson, N. (1963). *Information Theory and Coding*. New York, etc.: McGraw-Hill.

Aikman, D., Leake, M., & Nelson, B. (2014). Macroprudential instruments: a review. *Journal of Financial Stability, 11*, 65-84

Bachelier, L. (1900). Theory of Speculation, in Cootner P.H. ed. 1964. *The Random Character of Stock Market Prices. Cambridge, MA: MIT Press*

Bahrammirzaee, A. (2010). A comparative survey of artificial intelligence applications in finance: artificialneural networks, expert system and hybrid intelligent systems. *Neural Computing and Applications, 19*(8), 1165–1195.

Batagelj, V., Doreian, P., Ferligoj, A., & Kejzar, N. (2014). *Understanding large temporal networks and spatial networks: Exploration, pattern searching, visualization and network evolution*. Chichester: John Wiley & Sons.

Benedetto, F., Mastroeni, L., & Vellucci, P. (2021). Modeling the flow of information between financial time-series by an entropy-based approach. *Annals of Operations Research, 299*(1), 1235–1252.

Bianconi, G., Darst, R. K., Iacovacci, J., & Fortunato, S. (2014). Triadic closure as a basic generating mechanism of communities in complex networks. *Physical Review E, 90*(4), 042806.

Bonabeau, E. (2002). Agent-based modeling: Methods and techniques for simulating human systems. *Proceedings of the National Academy of Sciences, 99(3)*, 7280–7287.

Brock, W.A., and Hommes, C.H., (1997). Models of complexity in Economics and Finance, In: Hey, C. et al. (Eds.), *System Dynamics in Economic and Financial Models*, Chapter 1, 3-41, Wiley Publ.

Brooks, D. R., & Wiley, E. O. (1986). *Evolution as Entropy*. Chicago/London: University of Chicago Press.

Brunnermeier, M. K., & Oehmke, M. (2013). Bubbles, financial crises, and systemic risk. *Handbook of Economic and Finance, 2*, 1451-1556

Burt, R. S. (1982). *Toward a Structural Theory of Action*. New York, etc.: Academic Press.

Cagliarini, A., & Price, F. (2017). Exploring the Link between the Macroeconomic and Financial Cycles. *Research Papers in Economics*. tps://EconPapers.repec.org/RePEc:rba:rbaacv:acv2017-02

Carranza, A. R., Bejarano, J. L. R., & Bejarano, J. C. P. (2023). Shannon entropy to quantify complexity in the financial market (https://arxiv.org/pdf/2307.08666)

Chen, S., & Svirydzenka, K. (2021). Financial Cycles – Early Warning Indicators of Banking Crises? *IMF Working Papers*, *2021*(116), 1. https://doi.org/10.5089/9781513582306.001.A001

Chiarella, C., Dieci, R. and He, X. (2009). *Heterogeneity, market mechanisms, and asset price dynamics*. Elsevier.

Cho, P., & Kim, K. (2022). Global Collective Dynamics of Financial Market Efficiency Using Attention Entropy with Hierarchical Clustering. *Fractal and Fractional, 6* (10), 562.

Daubechies, I. (1992), *Ten Lectures on Wavelets.* Society for Industrial and Applied Mathematics: Philadelphia, PA, USA.

Dávila, J., & Korinek, A. (2017). Macroeconomic externalities of financial innovation. *The Review of Economic Studies, 85*, 62-97.

de Nooy, W., & Leydesdorff, L. (2015). The dynamics of triads in aggregated journal–journal citation relations: Specialty developments at the above-journal level. *Journal of Informetrics, 9*(3), 542-554. doi: 10.1016/j.joi.2015.04.005

Dhesi, G., & Ausloos, M. (2016). Modelling and measuring the irrational behaviour of agents in financial markets: Discovering the psychological soliton. *Chaos, Solitons & Fractals, 88*, 119-125.

Dubois D.M. (1998). Computing Anticipatory Systems with Incursion and Hyperincursion. In: Dubois D.M. (ed.). Computing Anticipatory Systems, CASYS-First International Conference (437, p. 3-29). Woodbury, NY: American Institute of Physics. https://doi.org/10.1063/1.56331

Etzkowitz, H. & Leydesdorff, L. (1995). The Triple Helix—University-Industry-Government Relations: A Laboratory for Knowledge-Based Economic Development. *EASST Review 14*(1), 14-19.

Etzkowitz, H. & Leydesdorff, L. (1998). The endless transition: A "triple helix" of university – industry – government relations, *Minerva 36*, 203-208.

Fama, E. F. (1970). Efficient capital markets: A review of theory and empirical work. *Journal of Finance, 25*(2), 383–417.

Gençay, R., & Gradojevic, N. (2017). The Tale of Two Financial Crises: An Entropic Perspective. *Entropy*, *19*(6), 244.

Granovetter, M. (1973). The Strength of Weak Ties. *American Journal of Sociology*, *78*, (6), 1360-80.

Hommes, C. (2006). *Heterogeneous agent models in economics and finance*, *Vol. 2*. Elsevier.

Ivanova, I. (2022). Information Exchange, Meaning and Redundancy Generation in Anticipatory Systems: Self-organization of Expectations - the case of Covid-19. *International journal of general systems, 51*(7), 675-690.

Ivanova, I. (2023). From the Triple Helix Model of Innovations to the Quantitative Theory of Meaning. *Triple Helix*, 196–204.

Ivanova, I. (2024). Communication of information in systems of heterogeneous agents and systems' dynamics, Quality & Quantity: International Journal of Methodology, Springer, *58*(6), 5377-5393.

Ivanova, I. and Leydesdorff, L. (2014a). Rotational Symmetry and the Transformation of Innovation Systems in a Triple Helix of University-Industry-Government Relations. *Technological Forecasting and Social Change 86,* 143-156. doi: 10.1007/s11192-014-1241-7

Ivanova, I. and Leydesdorff, L. (2014b). A simulation model of the Triple Helix of university-industry-government relations and the decomposition of the redundancy. *Scientometrics, 99*(3), 927-948.

Ivanova, I. and Rzadkowski, G. (2025). Triple Helix synergy and patent dynamics. Cross country comparison, *Quality & Quantity: International Journal of Methodology* (in press).

Ivanova, I., Rzadkowski, G. and Leydesdorff, L. (2025). Quantitative Theory of Meaning. Application to Financial Markets. EUR/USD case study. https://arxiv.org/abs/2410.06476

Ivanova, I. and Torday, J. (2025). Towards the Structure and Mechanisms of Complex Systems, the Approach of the Quantitative Theory of Meaning, *World Futures* (in preparation). https://arxiv.org/abs/2412.09007

Jenks, T. (2020). *Hyperwave Theory: The Rogue Waves of Financial Markets*. Archway Publishing.

Jordà, Ò., Schularick, M., & Taylor, A. M. (2013). When credit bites back: Leverage, business cycles, and crises. *Journal of Money, Credit and Banking, 45*, 3-28

Kaizoji, T. (2004). Intermittent chaos in a model of financial markets with heterogeneous agents. *Chaos, Solitons and Fractals*, *20*(2), 323-327.

Komorowski, P. (2017). The life cycle of a crisis - a model of the course of an economic collapse *Studia Ekonomiczne 325*, 95-108.

Korbel, J., Jiang, X., & Zheng, B. (2019). Transfer Entropy between Communities in Complex Financial Networks. *Entropy, 21*(11), 1124.

Krippendorff, K. (2009). Information of Interactions in Complex Systems. *International Journal of General Systems, 38*(6), 669-680.

LeRoy, S. F. and Porter, R. D. (1981). The present-value relation: Tests based on implied variance bounds. *Econometrica 49*, 555–574.

Leydesdorff, L. (2008). Configurational Information as Potentially Negative Entropy: The Triple Helix Model. *Entropy, 12*, 391-410.

Leydesdorff, L. (2021). The Evolutionary Dynamics of Discursive Knowledge: Communication-Theoretical Perspectives on an Empirical Philosophy of Science. In: *Qualitative and Quantitative Analysis of Scientific and Scholarly Communication* (Wolfgang Glänzel and Andrasz Schubert, eds.). Cham, Switzerland: Springer Nature.

Leydesdorff, L., Dolfsma, W., & Van der Panne, G. (2006). Measuring the knowledge base of an economy in terms of triple-helix relations among 'technology, organization, and territory'. *Research Policy, 35*(2), 181-199.

Leydesdorff, L., & Dubois, D. M. (2004). Anticipation in Social Systems: The Incursion and Communication of Meaning. *International Journal of Computing Anticipatory Systems, 15*, 203-216.

Leydesdorff, L., & Franse, S. (2009). The Communication of Meaning in Social Systems. *Systems Research and Behavioral Science, 26*(1), 109-117.

Leydesdorff, L., & Ivanova, I. A. 2014. Mutual Redundancies in Interhuman Communication Systems: Steps Toward a Calculus of Processing Meaning. *Journal of the Association for Information Science and Technology, 65*(2), 386-399.

Leydesdorff, L., Petersen, A., & Ivanova, I. (2017). The self-organization of meaning and the reflexive communication of information. *Social Science Information 56*(1), 4-27.

Lou, S., Lin, J. (2018). Rogue Waves in Nonintegrable KdV-Type Systems. *Chinese Physical Letters, 35*(5), 050202.

Lux, T. (1995). The socio-economic dynamics of speculative markets: interacting agents, chaos and the fat tails of return distributions, *Journal of Economic Behavior and Organization, 33*, 143-165.

Mandelbrot, B. B. (1997). *The variation of certain speculative prices*. Springer.

McGill, W.J. (1954). Multivariate information transmission. *Psychometrika, 19*, 97-116.

Mullainathan, S. and Spiess, J. (2017). Machine learning: An applied econometric approach. *Journal of Economic Perspective, 31(2)*, 87–106.

Murialdo, P., Ponta, L., & Carbone, A. F. (2020). Long-Range Dependence in Financial Markets:a Moving Average Cluster Entropy Approach. *Entropy, 22*(6), 634.

Papla, D., & Siedlecki, R. (2024). Entropy as a Tool for the Analysis of Stock Market Efficiency During Periods of Crisis. *Entropy*, *26*(12), 1079.

Petersen, A., Rotolo, D., & Leydesdorff, L. (2016). A Triple Helix Model of Medical Innovations: Supply, Demand, and Technological Capabilities in Terms of Medical Subject Headings. *Research Policy, 45*(3), 666-681.

Ponta, L., Murialdo, P., & Carbone, A. F. (2021). Information measure for long-range correlated time series: Quantifying horizon dependence in financial markets. *Physica A- Statistical Mechanics and Its Applications, 570*, 125777.

Reinhart, C. M., & Rogoff, K. S. (2009). *This time is different: Eight centuries of financial folly*. Princeton University Press

Rosen, R. (1985). *Anticipatory Systems: Philosophical, Mathematical and Methodological Foundations*. Oxford, etc.: Pergamon Press.

Rzadkowski G., and Figlia, G. (2021). Logistic wavelets and their application to model the spread of COVID-19 pandemic, *Appl. Sci.*, 8147.

Shannon C.E. (1948). A Mathematical Theory of Communication. *Bell System Technical Journal  27*, 379-423 and 623-656. https://doi.org/10.1002/j.1538-7305.1948.tb00917.x

Schmidhuber, J. (2015). Deep learning in neural networks: An overview. *Neural networks,* 61, 85-117.

Shiller, R.J. (1981). Do Stock Prices Move Too Much to be Justified by Subsequent Changes in Dividends? *American Economic Review, 71*(3), 421-436. https://doi.org/10.3386/w0456

Sinon, M.B.A., Mba, J.C. (2024). The analysis of diversification properties of stablecoins through the Shannon entropy measure. *Knowl Inf Syst* 66, 5501–5540.

Sokolov, A., Mostovoy, J., Parker, B., and Seco, L. (2020). Neural embeddings of financial time-series data. *The Journal of Financial Data Science, 2(4)*, 33–43.

Stefan, F., & Atman, A. (2015). Is there any connection between the network morphology and the fluctuations of the stock market index? Physica A: *Statistical Mechanics and Its Applications, (419)*, 630-641.

Sufi, A., & Taylor, A. M. (2021). Financial Crises: A Survey. University of Chicago, Becker Friedman Institute for Economics Working Paper No. 2021-97, Available at SSRN: https://ssrn.com/abstract=3906775

von Foerster, H. (1960). On self-organizing systems and their environments. In: *Self-Organizing Systems*, M.C. Yovits, S. Cameron eds., Pergamon, London, pp. 31–50.

Yan, Z. (2010). Financial Rogue Waves. *Communications in Theoretical Physics, 54*(4), 947-949.

Zha, Q., Kou, G., Zhang, H., Liang, H., Chen, X., Li, C.C, & Dong, Y. (2020). Opinion dynamics in finance and business: a literature review and research opportunities. *Financial Innovation, 6,* 44. https://doi.org/10.1186/s40854-020-00211-3